%08.12.2000
\magnification=1100
\headline{\hfill\folio}
\hsize=15.0 true cm
\vsize=22.0 true cm
\baselineskip=13 pt
\input psfig.sty 
\def\ref{\par\noindent\hangindent=1.0 true cm}
\nopagenumbers
\centerline{\bf RADIAL GRADIENTS AND METALLICITIES IN THE GALACTIC DISK
    \footnote \dag {\sevenrm Ionized Gaseous Nebulae, Rev. Mex. Astron. Astrof. SC, 
    in press}}
\bigskip
\bigskip
\centerline{WALTER J. MACIEL}
\bigskip
\centerline{IAG/USP, S\~ao Paulo, Brazil}
\bigskip
\centerline{\it maciel@iagusp.usp.br}
\centerline{\it http://www.iagusp.usp.br/$\sim$maciel}
\bigskip
\bigskip
\noindent 
\centerline{\bf ABSTRACT}
\bigskip
\vbox{\baselineskip 11 pt\noindent
Radial O/H abundance gradients derived from HII regions, 
hot stars and planetary nebulae are combined with [Fe/H] gradients from 
open cluster stars in order to derive an independent [O/Fe] $\times$ 
[Fe/H] relation for the galactic disk. A comparison of the obtained 
relation with recent observational data and theoretical models suggests 
that the [O/Fe] ratio is not higher than [O/Fe] $\simeq 0.4$, at least 
within the metallicity range of the considered samples.  }
\bigskip
\bigskip
%%%%%%%%%%%%%%%%%%%%%%%%%%%%%%%%%%%%%%%%%%%%%%%%%%%%%%%%%%%%%%%
\centerline{\bf 1. Introduction}
\medskip\noindent
%%%%%%%%%%%%%%%%%%%%%%%%%%%%%%%%%%%%%%%%%%%%%%%%%%%%%%%%%%%%%%%
The [O/Fe] $\times$ [Fe/H] ratio is one of the basic relationships in
the study of the chemical evolution of the Galaxy, as it links the
main metallicity indicators and stresses the different contribution
of Type II and Type~I supernovae. Recently, some discrepancy has been
observed between different sets of observational data and among
theoretical models themselves regarding the [O/Fe] abundance ratio.
Several of these results which are generally based on studies
of the [OI] forbidden line doublet at 6300 \AA, 6364 \AA\ in metal-poor 
giants lead to [O/Fe] $\simeq 0.5$ for [Fe/H] $\simeq -2$, showing a 
plateau in the [O/Fe] ratio at low metallicities (Barbuy \& Erdelyi-Mendes
1989, Edvardsson et al. 1993, Fulbright \& Kraft 1999, see 
Carretta et al. 2000 for additional 
references). Theoretical models such as those by Matteucci et al. (1999)
and Chiappini et al. (1997) are successful in reproducing these abundances 
for metallicities down to [Fe/H] $\simeq -2$. On the other hand, oxygen 
abundances from the OI infrared lines and some recent studies based 
on ultraviolet OH bands in metal-poor subdwarfs reach a much higher 
ratio, [O/Fe] $\simeq 1$ at low metallicities (Israeleian et al. 1998,
Boesgaard et al. 1999), which can be reproduced by theoretical models 
by Ramaty et al. (2000).

Some authors find that the slope of the [O/Fe] $\times$ [Fe/H]
relation is about $-0.30$ to $-0.40$, and essentially the same for the 
thin disk, thick disk and halo (Israelian et al. 1998, Boesgaard et al.
1999, Mishenina et al. 2000), so that any detailed investigation  should 
include both metal rich and metal poor objects belonging to these 
galactic subsystems. 

A contribution to the understanding of this problem can be obtained by
the analysis of radial abundance gradients in the galactic disk. Such
gradients can be observed both for the O/H ratio (and sometimes S/H, 
Ne/H and Ar/H as well) from HII regions, planetary nebulae and hot stars 
and also for the [Fe/H] metallicity, principally from open cluster stars 
(Maciel 1997, Maciel 2000). In this work, both sets of data 
are taken into account in order to derive an independent [O/Fe] $\times$ 
[Fe/H] relation appropriate to the galactic disk, roughly at metallicities 
[Fe/H] $\geq -1.5$. Such relation can be directly compared with recent 
observational data and with the predictions of detailed theoretical models, 
thus contributing to clarify the discrepancy in the [O/Fe] ratio
at subsolar metallicities.
\bigskip\bigskip
%%%%%%%%%%%%%%%%%%%%%%%%%%%%%%%%%%%%%%%%%%%%%%%%%%%%%%%%%%%%%%%
\centerline{\bf 2. Radial abundance gradients}
\medskip\noindent
%%%%%%%%%%%%%%%%%%%%%%%%%%%%%%%%%%%%%%%%%%%%%%%%%%%%%%%%%%%%%%%
\centerline{\underbar{2.1 The O/H gradient}} 
\medskip\noindent
%%%%%%%%%%%%%%%%%%%%%%%%%%%%%%%%%%%%%%%%%%%%%%%%%%%%%%%%%%%%%%%
The best determinations of the O/H radial gradient are made
on the basis of photoionized nebulae (HII regions and planetary
nebulae) and hot stars. Earlier work on these nebulae already
pointed to similar gradients, amounting roughly to
$d\log ({\rm O/H})/dR \simeq -0.06$ to $-0.07$ dex/kpc, as shown
by Shaver et al. (1983) and Fa\'undez-Abans \& Maciel (1986) 
for HII regions and planetary nebulae, respectively. 

More recent work has firmly established the presence of such
gradients, not only for the O/H ratio but also for other
elements such as S/H, Ne/H and Ar/H (see for example Maciel 1997
and Maciel 2000). HII region data have been discussed by 
Simpson et al. (1995), Esteban \& Peimbert (1995), V\'\i lchez \&
Esteban (1996) and Deharveng et al. (2000). Studies of planetary 
nebulae (Maciel \& K\"oppen 1994, Maciel \& Quireza 1999) go as 
far as to investigate possible space and time 
variations within the galactic disk, with important consequences 
for theoretical models.

Stellar data based on O and B stars apparently told a different
story. Several investigations up to 1997 presented controversial
results, with the general conclusion that no gradients were
present or were restricted to the inner parts of the disk 
(Fitzsimmons et al. 1990, Kaufer et al. 1994, Kilian-Montenbruck et 
al. 1994).  Since these stars were expected to have intermediate 
ages between HII regions and the planetary nebula central stars, 
it was difficult to conciliate these different sets of data. 
However, since the work of Smartt \& Rolleston (1997) and 
Gummersbach et al. (1998) these contradictions have been 
settled out and a clear gradient of the same order as the one 
derived from the photoionized nebulae has been obtained. This 
work was based on medium to high-resolution spectra of a relatively 
large sample of main sequence B stars in clusters and associations, 
spanning about 12 kpc in galactocentric distances. The O/H gradient 
from the above mentioned sources can be written as

$$\log ({\rm O/H}) + 12 = a + b \ R \ , \eqno(1)$$

\noindent
where $R$ is the galactocentric distance in kpc. Most determinations 
of the gradients assume  different distances of the LSR to the galactic 
center, ranging from 7 kpc to 8.5 kpc. For the sake of uniformity, we have
adopted a recent value $R_0 = 7.6$ kpc (Maciel 1993, Reid 1989),
so that the values of the gradients given here refer to this value.

The constants $a$ and $b$ have slightly different values depending 
on the nature of the objects considered, namely HII regions, planetary 
nebulae and hot stars. Generally speaking,  HII regions and hot 
stars have very similar gradients, while most planetary nebulae 
show a slightly flatter gradient. An average gradient of the former,
derived from the sources above and referring thus to the younger 
population, is characterized by the values $a = 9.34 \pm 0.14$ and  
$b = -0.070 \pm 0.014$ dex/kpc.
\bigskip\bigskip\noindent
%%%%%%%%%%%%%%%%%%%%%%%%%%%%%%%%%%%%%%%%%%%%%%%%%%%%%%%%%%%%%%%
\centerline{\underbar{2.2 The [Fe/H] gradient}} 
\medskip\noindent
%%%%%%%%%%%%%%%%%%%%%%%%%%%%%%%%%%%%%%%%%%%%%%%%%%%%%%%%%%%%%%%
A large amount of work has been done on the [Fe/H] gradient in the 
Galaxy, on the basis of data on open cluster stars (Janes 1995,
Friel 1995, Janes 2000). Most of these determinations indicate 
a steeper [Fe/H] gradient as compared with the O/H gradient derived 
from HII regions, hot stars and planetary nebulae. Results by
Friel, Janes and co-workers (Friel 1995, Janes 2000, Phelps 2000) 
suggest gradients of the order of $-0.07$ to $-0.09$ dex/kpc, 
which is in agreement with recent determinations by Twarog et al. 
(1998) and Carraro et al. (1998). The [Fe/H] gradient can be written as

$$[{\rm Fe/H}] = c + d \ R \eqno(2)$$

\noindent
and average values of the constants are $c = 0.50 \pm 0.05$ and 
$d = -0.085 \pm 0.008$ dex/kpc, again correcting for $R_0 = 7.6$ kpc.
\bigskip\bigskip\noindent
%%%%%%%%%%%%%%%%%%%%%%%%%%%%%%%%%%%%%%%%%%%%%%%%%%%%%%%%%%%%%%%
\centerline{\underbar{2.3 The time variation of the gradients}} 
\medskip\noindent
%%%%%%%%%%%%%%%%%%%%%%%%%%%%%%%%%%%%%%%%%%%%%%%%%%%%%%%%%%%%%%%
The time variation of the radial gradients is not well known, 
and in fact some theoretical models predict a time steepening of the 
gradients (Chiappini et al. 1997, Henssler 1999), while others predict 
just the opposite behaviour (Allen et al. 1998, Moll\'a et al. 1997). 
The main difficulty in approaching this problem is that objects with 
rather different ages should be considered, which introduces some 
difficulty in understanding the measured gradients. Probably the best 
group of objects to study this problem is the planetary nebulae, since 
these objects include both relatively old objects (the so-called 
type~III nebulae, see Peimbert 1978 and  Maciel 1989) and 
relatively young ones (the type~I and type~II nebulae).

Available data, as discussed by Maciel \& K\"oppen (1994) and 
Maciel \& Quireza (1999) point to a mild steepening of the gradients, 
so that the HII region gradient is slightly steeper than 
that of the planetary nebulae. Also, some steepening is 
observed comparing the gradients of  type III 
objects relative to the remaining, younger types, which supports this 
conclusion. On the other hand, the difference between the HII region 
and the stellar data is negligible, reflecting the similar ages of these 
objects or a constancy of the gradients at more recent times. In fact, 
recent models by Chiappini and collaborators (Chiappini et al. 1999
and private communication) predict a very slow steepening of the gradients 
during the last few Gyr, so that this assumption is probably correct. 
Therefore, we can consider the HII regions and open cluster stars as 
referring to essentially similar epochs, so that equations (1) 
and (2) and the given values of the constants $a$, $b$, $c$ and 
$d$ can be safely used as reflecting the present interstellar abundances of 
oxygen and iron in the galactic disk, within the assumed uncertainties.
\bigskip\bigskip
%%%%%%%%%%%%%%%%%%%%%%%%%%%%%%%%%%%%%%%%%%%%%%%%%%%%%%%%%%%%%%%
\centerline{\bf 3. Metallicities in the Galactic Disk}
\medskip\noindent
%%%%%%%%%%%%%%%%%%%%%%%%%%%%%%%%%%%%%%%%%%%%%%%%%%%%%%%%%%%%%%%
Adopting the usual definition of the abundance ratios relative to
the sun, $[{\rm X/Y}] = \log ({\rm X/Y}) - \log ({\rm X/Y})_\odot$, 
we have for the oxygen and iron abundances

$$[{\rm O/Fe}] = \alpha + \beta \ [{\rm Fe/H}]\ ,\eqno(3)$$

\noindent
where we have defined 

$$\alpha = a - {b \ c \over d} - [\log({\rm O/H})_\odot + 12]\eqno(4)$$

\noindent
and

$$\beta = {b \over d} - 1 \eqno(5)$$

\noindent
and have assumed that relations (1)  and (2) hold throughout 
the disk. Equivalently, we may write 
$[{\rm O/H}] = \alpha + (\beta + 1) \ [{\rm Fe/H}]$. Another interesting
relation can be written as

$$[{\rm Fe/H}] = \gamma + \delta \ [\log({\rm O/H}) + 12]\eqno(6)$$

\noindent
with 

$$\gamma = c - {a \, d\over b} \eqno(7)$$

\noindent
and

$$\delta ={ d \over b} \ . \eqno(8)$$

\noindent
It is easy to show 
that $\delta = 1/(1 + \beta)$ and $-\gamma/\delta = \alpha  +  
[\log({\rm O/H})_\odot + 12]$, so that the parameters $\gamma$ and 
$\delta$ are not independent.
\bigskip\bigskip
%%%%%%%%%%%%%%%%%%%%%%%%%%%%%%%%%%%%%%%%%%%%%%%%%%%%%%%%%%%%%%%
\centerline{\bf 4. Results and discussion}
\medskip\noindent
%%%%%%%%%%%%%%%%%%%%%%%%%%%%%%%%%%%%%%%%%%%%%%%%%%%%%%%%%%%%%%%
In view of the previous discussion, adopting the given values of the 
constants $a$, $b$ and $c$, we obtain the following parameters:
$\alpha = 0.098$, $\beta = -0.176$, $\gamma = -10.841$ and
$\delta = 1.214$, where we have used the solar abundances 
$\log$(O/H)$_\odot$ + 12 = 8.83 (Grevesse \& Sauval 1998).

The main results are shown in Figure~1, which includes two
panels, the first showing the usual [O/Fe] $\times$ [Fe/H] relationship 
and the second showing the [Fe/H] $\times$ $\log$(O/H) + 12 relationship. 
The solid lines show the predicted \lq\lq theoretical \rq\rq\ relationships 
taking into account the observed radial gradients. The dotted lines show 
results of theoretical models by Matteucci et al. (1999) and the dot-and-dashed 
lines represent models by Ramaty et al. (2000). Data from Matteucci et
al. (1999) follow models by Chiappini et al. (1997), and are 
representative of models predicting a [O/Fe] plateau for metallicities 
under solar, without a significant increase in the [O/Fe] ratio above
0.5 dex for [Fe/H] $\leq -2$. On the other hand, models by 
Ramaty et al. (2000) have been selected to display higher [O/Fe] ratios at 
lower metallicities, adopting supernova yields from Woosley \&
Weaver (1995) and finite Fe and O mixing delay times. Figure~1 
also includes some representative observational data by Barbuy \&
Erdelyi-Mendes (1989), asterisks; Boesgaard et al. (1999), filled 
squares; Edvardsson et al. (1993), crosses; Israelian et al. 
(1998), solid dots; Spiesman \& Wallerstein (1991), open circles; 
Spite \& Spite (1991), plus signs; Takeda et al. (2000), empty squares; 
Mishenina et al. (2000), open stars, and Cavallo et al. (1997), filled stars, 
as recomputed by Mishenina et al. (2000).

\bigskip
%----------------------------------------------------------------
\centerline{\psfig{figure=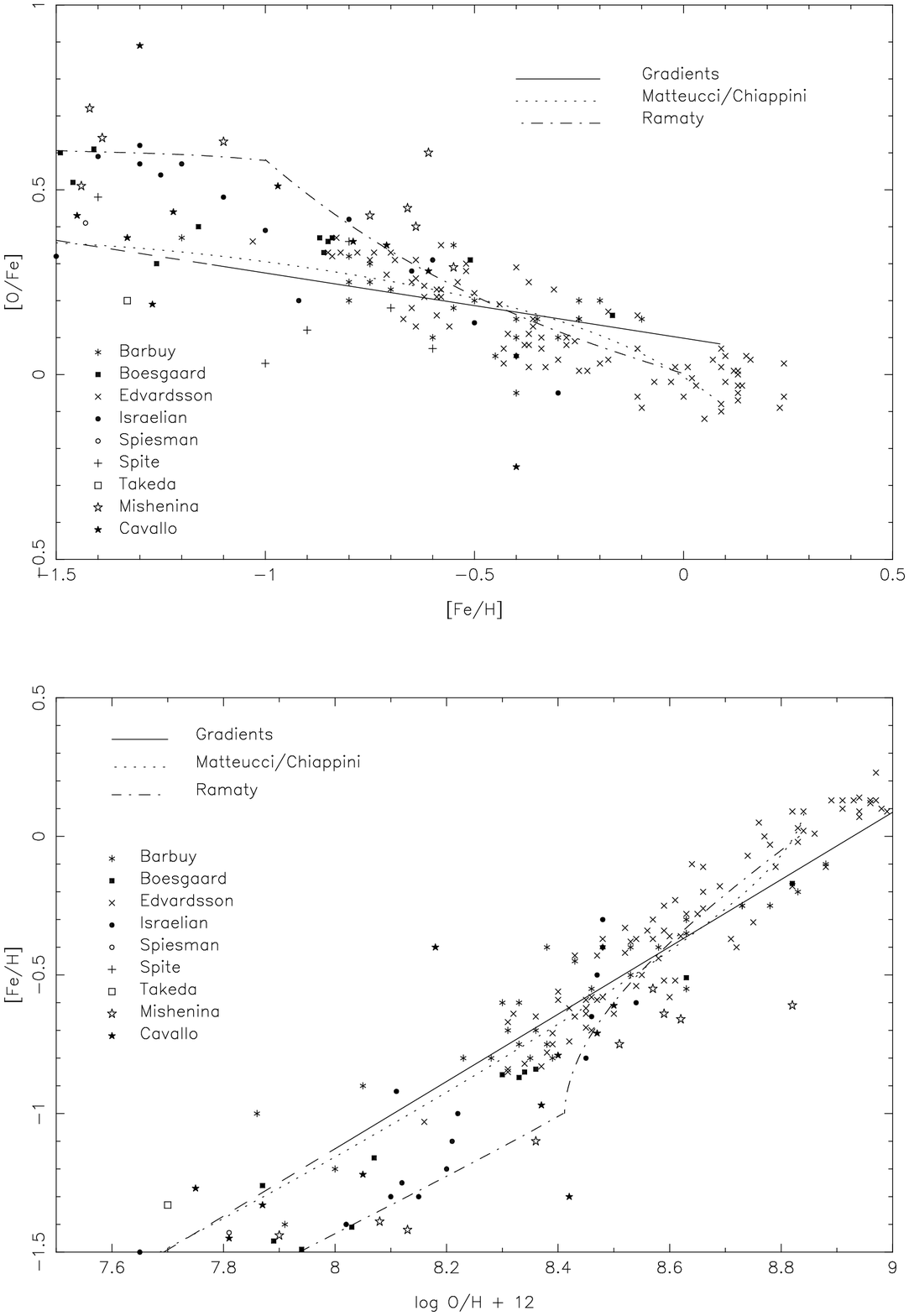,height=16.0 cm,angle=0} }
\centerline{\sevenrm \baselineskip 12 pt Figure 1 -  
    Metallicities in the galactic disk.}
%----------------------------------------------------------------
\bigskip

It can be seen that for metallicities close to and slightly lower than
the solar value, all observational data and models show a reasonable 
agreement, while for lower metallicities the spread is considerably
larger. Part of this scatter may be due to the use of different
scales of stellar parameters such as effective temperatures,
gravities and metallicities, to the adoption of different atomic
parameters or the neglecting of NLTE effects. However, the spread
is large enough so that two different regimes can be distinguished,
namely, \lq\lq low [O/Fe]\rq\rq\ and \lq\lq high [O/Fe]\rq\rq\ at
low metallicities. The data by Boesgaard et al. (1999) and 
Israelian et al. (1998) indicate higher [O/Fe] ratios, 
closer to the models by Ramaty et al. (2000).
The remaining data do not show such large [O/Fe] abundances,
in agreement with the models by Matteucci et al. (1999) and 
Chiappini et al. (1997). 

It is clear that the gradient data support the 
lower [O/Fe] abundances predicted by the latter, at least for 
metallicities larger than [Fe/H] $\simeq -1.5$, which are appropriate 
for the oldest populations of the disk. The gradients themselves cover 
a smaller fraction of the metallicity range, as shown by the solid 
lines in figure~1, and already in this region it is apparent a  
better agreement with the model predictions of Matteucci et al. (1999). 
Extrapolating the solid lines towards lower metallicities, 
as shown by the broken lines, we obtain an upper limit for the 
[O/Fe] ratio of 0.4 dex, which is also closer to the low [O/Fe] regime.

Therefore, our data is consistent with a maximum [O/Fe] $\simeq 0.4$ 
for the galactic disk, at least for metallicities as low as 
[Fe/H] $\simeq -1.5$, which are appropriate to the region where the 
abundance gradients are observed. This is also supported by 
recent observations of infrared OH lines in metal-poor stars
(Mel\'endez et al. 2000) and by a recent analysis by Carretta et
al. (2000), based on data for a selected sample of 19 stars. Their results 
support a moderate increase in the [O/Fe] ratio, pointing out 
that the higher [O/Fe] obtained by Boesgaard et al. (1999) derive mostly 
from OH abundances, which suffer from a  difficulty in the location 
of the continuum levels, so that these abundances are probably not 
reliable. Also, the analysis of Fulbright \& Kraft (1999) of some of the 
stars studied by Israelian et al. (1998) concludes that the [O/Fe] ratio 
is lower than derived earlier, and that it is premature that the oxygen 
abundances of metal-poor stars should be increased. Of course, 
the simplicity of the linear gradients does not allow to
predict any change of slope or the presence of a plateau, so that
the agreement between the solid lines of Figure~1, the
metallicity data and theoretical models is only approximate. However,
it is clear that no [O/Fe] ratio higher than about 0.4 dex 
can be expected for the metallicities considered. Furthermore, a small 
upward trend in [O/Fe] is also observed by Carretta et al. (2000), leading 
to maximum ratios close to the results shown in Figure~1.

The second panel in Figure~1 is particularly useful for photoionized 
nebulae such as HII regions and planetary nebulae, for which the [Fe/H] abundance
cannot be usually obtained directly. In this case, an average relation such
as shown by the solid straight line could be used to estimate the 
expected [Fe/H] for a given oxygen abundance. Also, if a determination
of [Fe/H] is available, this line could be used to estimate the
amount of iron that is condensed in solid grains.

\bigskip
%----------------------------------------------------------------
\centerline{\psfig{figure=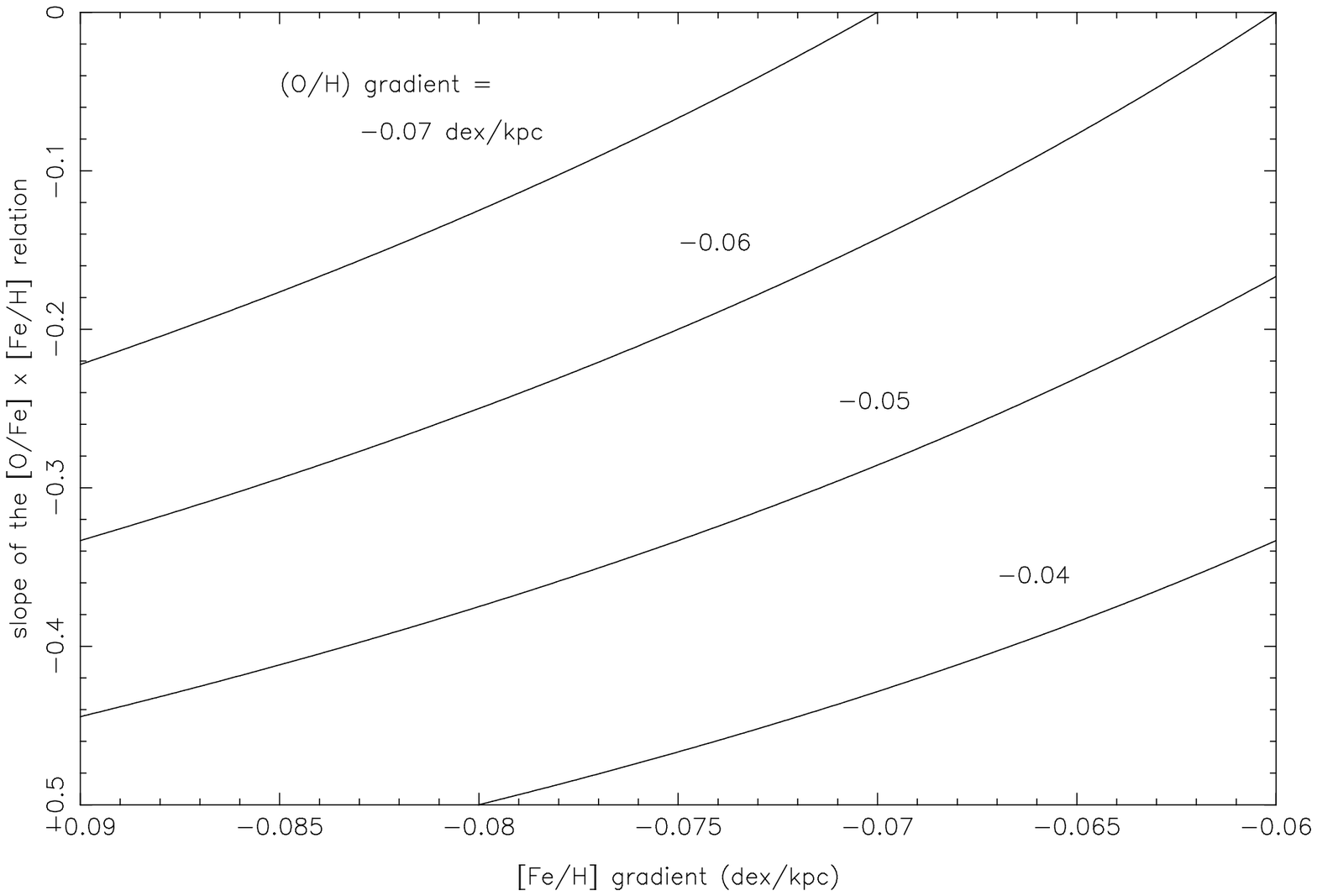,height=8.0 cm,angle=0} }
\centerline{\sevenrm \baselineskip 12 pt Figure 2 -  
    Slope of the [O/Fe] $\times$ [Fe/H] relation in the 
    galactic disk.}
%----------------------------------------------------------------
\bigskip

The straight lines in Figure~1 should be considered
as average relations, as no effort has been made to fit either
the solar or the remaining data. For example, these lines
could be slightly displaced vertically if we take into account
the uncertainties in the intercepts $a$ and $c$ of equations 
(1) and (2), respectively. The effect of the 
uncertainties in the determination of the gradients
can be observed in Figure~2, where we show the slope
$\beta$ of the [O/Fe] $\times$ [Fe/H] relation as a function of the
[Fe/H] gradient for different values of the O/H gradient. It can be
seen that the average slope is about $\beta \simeq -0.2$, and that
steeper slopes would require either a very low O/H gradient,
assuming that the [Fe/H] gradient is $-0.085$ dex/kpc, or
a very steep [Fe/H] gradient, assuming that the O/H gradient
is  $-0.07$ dex/kpc. Since there is no observational
evidence for these possibilities, it can be concluded that
average slopes steeper than $\beta \simeq -0.2$ are not supported
by the present data.
\bigskip
\centerline{\bf Acknowledgements}
\medskip\noindent
I thank B. X. Santiago and C. Chiappini 
for some fruitful discussions. This work has been partially 
supported by CNPq, FAPESP and CAPES.
\bigskip
\centerline{\bf REFERENCES}
\medskip
\ref
Allen, C.,  Carigi, L. \& Peimbert, M. 1998, ApJ {494}, 247
\ref
{Barbuy}, B. \& Erdelyi-Mendes, M. 1989, A\&A {214}, 239 
\ref
Boesgaard, A.~M., King, J.~R., Deliyannis, C.~P. \& Vogt, S.~S. 
   1999, AJ {117}, 492 
\ref
Carraro, G., Ng, Y. K. \& Portinari, L. 1998, MNRAS {296}, 1045
\ref
Carretta, E., Gratton, R.~G. \& Sneden, C. 2000, A\&A {356}, 238
\ref
Cavallo, R.~M., Pilachowski, C.~A. \& Rebolo, R. 1997, 
    PASP {109}, 226
\ref
Chiappini, C., Matteucci, F. \& Gratton, R. 1997, ApJ {477}, 765
\ref
Chiappini, C., Matteucci, F., Beers, T.~C. \& Nomoto, K. 1999, 
   ApJ {515}, 226 
\ref
Deharveng, L., Pe\~na, M., Caplan, J. \& Costero, R. 2000, MNRAS {311}, 329
\ref
Edvardsson, B., Andersen, J., Gustafsson, B., Lambert, D.~L., 
   Nissen, P.~E. \& Tomkin, J. 1993, A\&A  {275}, 101 
\ref
Esteban, C. \& Peimbert, M. 1995, RMAA SC {3}, 133 
\ref
Fa\'undez-Abans, M. \& Maciel, W.~J. 1986 A\&A {158}, 228
\ref
Fitzsimmons, A., Brown, P.~T.~F., Dufton, P.~L. \& Lennon, D.~J. 
    1990, A\&A {232}, 437
\ref
Friel, E.~D. 1995, Ann. Rev. A\&Ap {33}, 381
\ref
Fulbright, J.~P. \& Kraft, R.P. 1999, ApJ {118}, 527
\ref
Grevesse, N. \& Sauval, A.~J. 1998, Space Sci. Rev. {85}, 161
\ref
Gummersbach, C.~A., Kaufer, A., Sch\"afer, D.~R., Szeifert, T. \& 
    Wolf, B. 1998, A\&A {338}, 881
\ref
Henssler, G. 1999, Ap\&SS {265}, 397
\ref
Israelian, G., Garc\'\i a L\'opez, R.~J. \& Rebolo, R. 1998, 
   ApJ {507}, 805 
\ref
Janes, K. 1995, In: A. Alfaro \&  A.~J. Delgado (eds.): 
    The Formation of the Milky Way, Cambridge, 144
\ref
Janes, K. 2000, In: F. Giovannelli \& F. Matteucci (eds.):
    Chemical Evolution of the Milky Way: Stars versus Clusters, 
    Kluwer (in press)
\ref
Kaufer, A., Szeifert, Th., Krenzin, R., Baschek, B. \& Wolf, B.
    1994, A\&A {289}, 740
\ref
Kilian-Montenbruck, J., Gehren, T. \&  Nissen, P.~E. 1994, 
    A\&A {291}, 757
\ref
Maciel, W.~J. 1989, In: S. Torres-Peimbert (ed.):
    IAU Symp. 131, Kluwer, Dordrecht, 73 
\ref
Maciel, W.~J. 1993, Ap\&SS {206}, 285
\ref
Maciel, W.J. 1997, In: H.~J. Habing \& H.~J.~G.~L.~M. 
    Lamers (eds.): IAU Symp. 180, Kluwer, Dordrecht, 397
\ref
Maciel, W.J. 2000, In: F. Giovannelli \& F. Matteucci (eds.):
    Chemical Evolution of the Milky Way: Stars versus Clusters, 
    Kluwer (in press)
\ref
Maciel, W.~J. \& K\"oppen, J. 1994, A\&A {282}, 436
\ref
Maciel, W.~J., Quireza, C. 1999, A\&A  {345}, 629
\ref
Matteucci, F., Romano, D. \& Molaro, P. 1999, A\&A {341}, 458
\ref
Mel\'endez, J., Barbuy, B. \& Spite, F. 2000, ApJ (in press)
\ref
Mishenina, T.~V., Korotin, S.~A., Klochkova, V.~G. \& 
    Panchuk, V.~E. 2000, A\&A {353}, 978
\ref
Moll\'a, M., Ferrini, F. \& D\'\i az, A.~I. 1997, 
    ApJ {475},  519
\ref
Peimbert, M. 1978, In: Y. Terzian (ed.): IAU Symp. 76, Reidel,
    Dordrecht, 233
\ref
Phelps, R. 2000, In: F. Giovannelli \& F. Matteucci (eds.):
    Chemical Evolution of the Milky Way: Stars versus Clusters, 
    Kluwer (in press) 
\ref
Ramaty, R., Scully, S.~T., Lingenfelter, R.~E., Kozlovsky, B. 
   2000, ApJ  {534}, 747
\ref
Reid, M.~J. 1989, In: M. Morris (ed.): The Center of the Galaxy, 
    Reidel, 37
\ref
Shaver, P.~A., McGee, R.~X., Newton, L.~M., Danks, A.~C. \& 
    Pottasch, S.~R. 1983, MNRAS {204}, 53
\ref
Simpson, J.~P., Colgan, S.~W.~J., Rubin, R.~H., Erickson, E.~F. 
    \& Haas, M.~R. 1995, ApJ {444}, 721
\ref
Smartt, S.~J. \& Rolleston, W.~R.~J. 1997, ApJ {481}, L47
\ref
Spiesman, W.~J. \& Wallerstein, G. 1991, AJ {102}, 1790
\ref
Spite, M. \& Spite, F. 1991, A\&A {252}, 689
\ref
Takeda, Y., Takada-Hidai, M., Sato, S., Sargent, W.~L.~W., 
    Lu, L., Barlow, T.~A. \& Jugaku, J. 2000, ApJ (in press)
\ref
Twarog, B.~A., Ashman, K.~M. \& Anthony-Twarog, B.~J. 1997, 
    AJ {114}, 2556  
\ref
V\'\i lchez, J.~M. \& Esteban, C. 1996, MNRAS {280}, 720
\ref
Woosley, S.~E. \& Weaver, T.~A. 1995, ApJS {101}, 181
\ref
\bye